\documentclass[2014, preprint, 1p,times,12pt,sort&compress]{elsarticle}

\usepackage{amssymb}

\usepackage{mathptmx}  		
\usepackage{amsmath}			

\usepackage{changepage}

\usepackage{graphicx}
\usepackage[section]{placeins}
\usepackage[normalsize]{caption}
\usepackage{subcaption}
\usepackage{pgfplots}					
\usepackage{psfrag}

\usepackage{natbib}
\journal{Journal of Computational Physics}

\begin{document}

\begin{frontmatter}



\title{A New Multi-Scale Structure Finding Algorithm to
Identify Cosmological Structure}


\author[1]{Ali Snedden\corref{cor1}} \ead[url]{asnedden@nd.edu}
\author[1]{Lara Arielle Phillips} \author[1]{Grant J. Mathews} \author[1]{Jared Coughlin} \author[1]{In-Saeng Suh}  \author[1]{Aparna Bhattacharya}

\address[1]{Center for Astrophysics\\ Department of Physics \\225 Nieuwland Science Hall\\ University of Notre Dame\\Notre Dame, IN 46556}
\cortext[cor1]{Corresponding author. Tel.: 574-631-3583}

\begin{abstract}
We introduce a new self-consistent structure finding algorithm that parses large
scale cosmological structure into clusters, filaments and voids. This structure
finding algorithm probes the structure at multiple scales and classifies
the appropriate regions with the most probable structure type and size. We
show that it reproduces key observational results, including the baryon fraction of ICM, and average temperatures and densities for the different structures.

\end{abstract}


\end{frontmatter}


\section{Introduction}
\label{sec:intro}
The Universe began as a relatively homogenous plasma with very small initial
cosmic density and entropy perturbations (as evidenced in the power spectrum of
the cosmic microwave background radiation). These perturbations evolved
through non-linear gravitational processes into the galaxies that are observed today.
On the largest of scales, these galaxies are components of even larger
structures seeded by the initial perturbations.

Numerous large galaxy redshift surveys \cite{joeveer78, delapparent86, geller89, colless01, gott05} have revealed filaments and walls connecting clusters together with vast voids in between.  This large scale structure (LSS) has been dubbed \citep{bond96}  the ``cosmic web."  The features within the LSS can be from several to hundreds of megaparsecs in scale.  Some of the most well known observed features include the Bo\"{o}tes void, the Great Wall and the Sloan Great Wall \citep{kirshner81, geller89, gott05}.  

Moreover, it has been shown \citep{abell65, oemler74, dressler80, grogin99, grogin00, rojas04, aragon07b, hahn07, paz08, zhang09, mo10, kreckel11} that galaxy properties can depend on their environment. Galaxies in clusters tend to be redder and elliptical when compared to those in the field \citep{abell65, oemler74, dressler80, mo10}.  Conversely, galaxies in voids tend to be spiral and bluer \citep{grogin99, grogin00, rojas04, kreckel11}.  From simulations, it has also been noted  \citep{aragon07b, hahn07, paz08, zhang09} that the spin of the dark matter halo associated with a galaxy is dependent on orientation within a filament.  It is clear that the large scale environment must play a critical role in the formation and evolution of galaxies.  

It is a challenge, however, to find an optimum algorithm that can successfully identify clusters, filaments and voids in large sets of observational or numerical data. Many different methods \citep{moody83, hoyle04, rojas04, ceccarelli06, aragon07a,  hahn07, bond08, colberg08, tinker08, foster09, bond10, kreckel11, platen11, tempel13} have been proposed for identifying structures and their associated galaxies within large galaxy redshift surveys and cosmological simulations. Most algorithms \citep{moody83, hoyle04, rojas04, ceccarelli06, bond08, colberg08, tinker08, foster09, bond10, kreckel11, platen11, tempel13} only follow one particular type of structure. Moreover, there are  significant ambiguities in the definitions of what exactly constitutes  a filament or void (although clusters are fairly well agreed upon).  Hence, a comparison of structures found by different algorithms remains a challenging task \citep{colberg08}.

This lack of consistency makes it difficult to use different algorithms to follow each individual type of structure (e.g. clusters, filaments and voids). Only a few algorithms  \citep{aragon07a, hahn07} follow multiple types of structure. Here we present a new structure finding algorithm adapted to solve this problem.

The algorithm described here is similar to those presented by Refs. \citep{hahn07,aragon07a, bond08}.  It utilizes the density curvature information provided by the eigenvalues of the Hessian matrix to find structures. However, unlike Refs. \cite{aragon07a, bond08}, the structure finder presented here provides a self-consistent way for dealing with clusters, filaments \emph{and} voids.

This new algorithm also has the advantage of being computationally more efficient than the structure finders of Refs. \citep{aragon07a, hahn07}.  We use the density obtained directly from the simulation output, without the need to compute either a Delaunay Tessellation Field Estimator \cite{aragon07a} or the gravitational potential and particle trajectories \cite{hahn07}.

\section{Methods}
\label{sec:methods}
The structure finding algorithm described here uses $2^{nd}$ order curvature information from the density distribution to parse the large scale structure.  It is based on a vasculature segmentation algorithm originally developed in Ref.~\cite{descoteaux04} for application to medical imaging. The algorithm uses the local curvature information to filter \citep{frangi98, descoteaux04} the filaments from other structures.  To get this curvature information one can Taylor expand the image, ${I(\bf x, \sigma)}$, about the local position ${{\bf x_{0}}}$ at  some characteristic scale, $\sigma$.  The resulting expansion is 
\begin{equation}
I({\bf x_0} + \delta {\bf x}, \sigma) \approx I({\bf x_0}, \sigma) + \nabla I({\bf x_0}, \sigma)_i   \delta {\bf x}^i + {\frac{1}{2}} H({\bf x_0}, \sigma)_{ij}   \delta {\bf x}^i  \delta {\bf x}^j  + ....
\label{eq:m40}
\end{equation}
where $\nabla I({\bf x_0}, \sigma)_i$ is the gradient of the image, and $H({\bf x_0}, \sigma)_{ij}$ is the 3 $\times$ 3 Hessian matrix  comprised of the second derivatives of the image and thereby characterizing the image curvature. To calculate the spatial derivatives, the image is convolved \citep{frangi98, florack92} with a kernel composed of the derivatives of a Gaussian function along the desired dimension .  Thus, 
\begin{equation}
H({\bf x_0, \sigma})_{ij} = \int \nabla I({\bf x_0}, \sigma)_i * \frac{\partial G({\bf x - x_0}, \sigma)}{\partial {x}_j} dx_j~~,
\label{eq:m41}
\end{equation}
where
\begin{equation}
\nabla I({\bf x_0}, \sigma)_{i} = \int I({\bf x}, \sigma) * \frac{\partial G({\bf x - x_0}, \sigma)}{\partial {x}_i} dx_i~~
\label{eq:m42}
\end{equation}
and 
\begin{equation}
G({\bf x - x_0}, \sigma) = {\frac {1}{\sqrt{2\pi\sigma^2}}}e^{-\frac{({\bf x}-{\bf x}_0)^2}{2\sigma^2}}.
\label{eq:m43}
\end{equation}
The characteristic Gaussian scale, $\sigma$, can then be modified to effectively probe
filaments of different sizes by changing the width of the Gaussian kernel. For
instance, filaments that are smaller in the radial dimension will be best probed
by narrower Gaussian kernels because they change more rapidly over a small
distance. The opposite holds true for larger filaments.

The Hessian matrix is multiplied by the characteristic scale to account for
the smoothing effect that large scales have on the matrix values. The eigenvalues
and eigenvectors are then calculated. The resulting eigenvectors yield three
orthogonal axes with the eigenvalues giving the relative spatial curvature along
each axis \citep{frangi98}.  These eigenvalues are ordered by their relative magnitudes as
\begin{equation}
|\lambda_1| \leq |\lambda_2| \leq |\lambda_3|.
\label{eq:m44}
\end{equation}   
The relative ratios of the eigenvalues is what determines  the structural information.  For example, if $|\lambda_1| << |\lambda_2|$ and $|\lambda_2| \approx |\lambda_3|$, then, one dimension has a relatively small curvature, while two dimensions have much higher curvature. This signature is indicative of a filament. The eigenvector associated with $|\lambda_1|$ must then be pointing down the length of the filament, with the other two eigenvectors perpendicular to it.  


Here, we generalize this algorithm to other structures such as sheets, clusters and voids (see Table 1). To parse the different structures (e.g. clusters, sheets and filaments) the spatial derivatives are taken with respect to the three dimensional density image. The relative ratios of the eigenvalues can then be used to generate a unique structure measure for each type of structure. Following the logic from Refs.~\cite{frangi98, descoteaux04}, structure measures for the clusters, sheets, and filaments can be represented by 3 quantities:
\begin{equation}
V_c = (1- e^{-2 |\lambda_1/\lambda_3|^2})*(1 - e^{-2 |F_{norm}|^2})~~,
\label{eq:m45}
\end{equation}
\begin{equation}
V_s = (1- e^{-2 |\lambda_2/\lambda_3|^2})*(1 - e^{-2 |F_{norm}|^2})~~,
\label{eq:m46}
\end{equation}
\begin{equation}
V_f = (1- e^{-2 |\lambda_1/\lambda_2|^2})*(1 - e^{-2 |F_{norm}|^2})~~,
\label{eq:m47}
\end{equation}
respectively, where $F_{norm}$ is the Frobenius norm, 
\begin{equation}
F_{norm} = ({\lambda_1}^2 + {\lambda_2}^2 + {\lambda_3}^2)^{\frac{1}{2}}. 
\label{eq:m48}
\end{equation}
The Frobenius norm term acts to wash out any spurious structure signal by
suppressing random eigenvalue fluctuations from the noise. Building upon this
logic, we introduced a new structure measure that is useful for voids
\begin{equation}
V_v = (1- e^{-2 |\lambda_1/\lambda_3|^2})*(e^{-2 |F_{norm}|^2}).
\label{eq:m49}
\end{equation}
Here, the role of the Frobenius norm term has changed. Instead of decreasing
the signal from spurious noise, it enhances the signal from the background.
The reasoning is that regions of low density will be dominated by the spurious
background noise and not by the actual physical structure. This provides a
self-consistent method with which to investigate simultaneously all four possible
morphological large scale structures observed in the Universe. 

Here, an additional condition is imposed on the structure measures. For clusters
and filaments, the structure measure was set to 0 if the volume element is in
an under-dense region. For the voids, the structure measure was set to 0 if the
volume element is in an over-dense region. This avoids confusing the identification
of structures due to smaller scale features.

For the purposes of this study, however, the sheet-like structures are not explicitly
traced due to the difficulty in distinguishing them from the filaments.

\begin{table}[h]
	\centering
	\begin{tabular}{| l | c | c | r|}
	\hline
		$\lambda_1$ & $\lambda_2$ & $\lambda_3$ & structure type\\
		\hline 
		$-$                 &     $-$             &  $-$               & cluster\\
		$\approx$ 0   &     $-$             &  $-$               & filament\\
		$\approx$ 0   &   $\approx$ 0  &  $-$               & sheet\\
		$+$                 &     $+$             &   $+$              & void\\
		\hline
		\end{tabular}
\caption{The signs of the eigenvalues for the characterization of different structures. The
clusters, sheets and filaments are negative due to the structure consisting
of large density values in a sea of small values. In the void case, the structure
being sought is the opposite of clusters, hence the opposite sign.}
\label{tab:m1}
\end{table}

\section{Results}
\label{sec:results}
In Section \ref{sec:millennium_run}, the structure finding algorithm was run on a sub-volume of the dark-matter only Millennium Simulation \cite{springel05, croton06, colberg08}.  This allows us to test the validity of our new void finding algorithm by benchmarking with respect to Ref.~\cite{colberg08}.  In Section \ref{sec:gadget_simulation}, we explore the validity of the entire algorithm in finding clusters, filaments and voids in a cosmological simulation [including supernova feedback] and comparing it to observed results. 

\subsection{Millennium Simulation}
\label{sec:millennium_run}
A (60 $h^{-1}$Mpc)$^{3}$  sub-volume containing 12,528,667 dark matter particles was extracted from the Millennium Simulation \cite{springel05} for analysis.  The Millennium Simulation is a dark matter simulation containing 2160$^{3}$ particles in a (500 $h^{-1}$Mpc)$^{3}$ volume with a particle mass resolution of 8.6 $\times$ 10$^{8}$ $h^{-1}$ M$_{\bigodot}$.   A $\Lambda$CDM cosmology is used with $\Omega_{m}$ = 0.25, $\Omega_{\Lambda}$ = 0.75 and $h$ = 0.73.  The region was overall underdense, with a density at 72\% of the mean cosmic density.  The associated 17,604 galaxies were also extracted from the semi-analytic galaxy catalogue generated by Ref.~\cite{croton06}.  This was the same volume used by Aspen-Amsterdam Void Finder Comparison Project \citep{colberg08}.

The structure finder was used to find voids within this region and then compared
to the void finders described in Ref.~\cite{colberg08}. This provides a benchmark from
which to compare the validity of the void finding algorithm.

To convert the particle data to grid data, the dark matter particles were binned
onto a 300$^3$ grid mapped to the sub-volume. The data was then smoothed using a
Gaussian kernel with a standard deviation equal to 0.2 $h^{-1}$Mpc  (the length of the side of 1 voxel).  This slight smoothing was necessary to ensure the continuity of the derivatives.  Without it, the data would be pixelated and have discontinuous derivatives.

\begin{figure}[h]
	\begin{center}
	\begin{subfigure}{.5 \textwidth}
		\centering
		\includegraphics[width=2.0in]{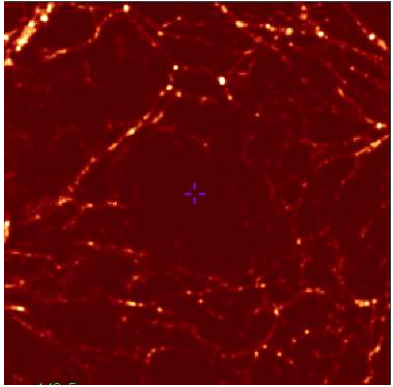}
		\caption{Dark matter distribution.}
		\label{fig:mnc_void.a}
	\end{subfigure}%
	\begin{subfigure}{.5 \textwidth}
		\centering
		\includegraphics[width=2.0in]{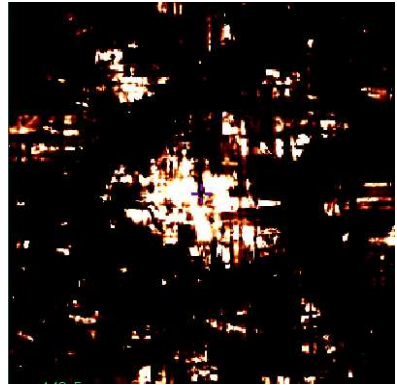}
		\caption{Void regions highlighted.}
		\label{fig:mnc_void.b}
	\end{subfigure}
	\end{center}
\caption{Aspen-Amsterdam Void}{Illustrates a one voxel thick slice through a selection of a void region within a sub-volume of the Millennium Simulation.  Fig.~\ref{fig:mnc_void.a} illustrates the smoothed dark matter distribution in log space.}
\label{fig:mnc_void}
\end{figure}

\begin{figure}[h]
	\begin{center}
	\psfrag{Mpc/h}[c][c][0.5]{$h^{-1}$Mpc}
	\includegraphics[width=4.0in]{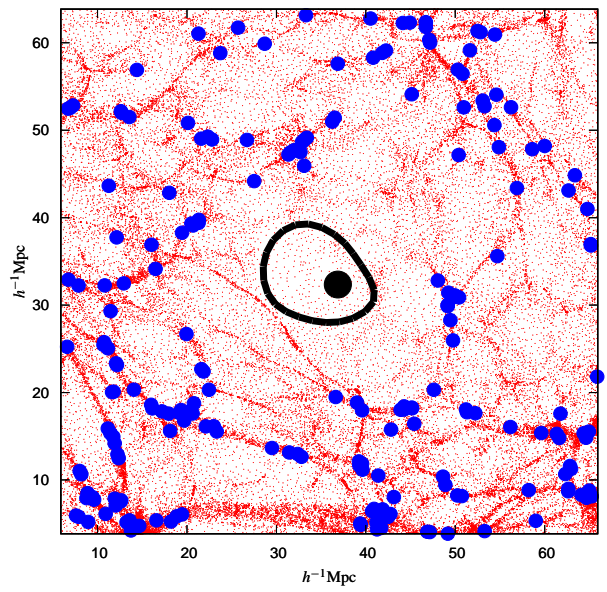}
	\end{center}
	\caption{Void Outlined in Particle Data.}{Illustrates a 1.0 $h^{-1}$Mpc thick (equal to 5 voxels) slice through the void region within a sub-volume of the Millennium Simulation.  Red dots represent individual dark matter, blue points represent galaxies from the semi-analytic catalogue, the big black dot represents the geometric center of the void and the contour outlines the connected, central void region}
\label{fig:void_contour}
\end{figure}

The structure finder was run on this data using  10 characteristic scales ($\sigma =$ 10, 13, 16, 19, 22, 25, 28, 31, 34 and 37 voxels in size).  Fig.~\ref{fig:mnc_void.a} provides a 1 voxel thick slice view of the entire volume and Fig. 4b shows the void region
highlighted. It is qualitatively evident that the void finder selects the central most
under-dense region.  Fig.~\ref{fig:void_contour} provides a 5 voxel thick view of the void overlaid on the dark matter and galaxy distribution.  This compares favorably to Fig. 1 of Ref.~\cite{colberg08}, and illustrates that the central void has been correctly identified.  

\begin{figure}[h]
	\begin{center}
	\psfrag{X}[tc][c][1]{Radius $h^{-1}$Mpc}
	\psfrag{Y}[c][c][1]{$\frac{\rho}{\bar{\rho}}$ - 1}
	\includegraphics[width=4.0in]{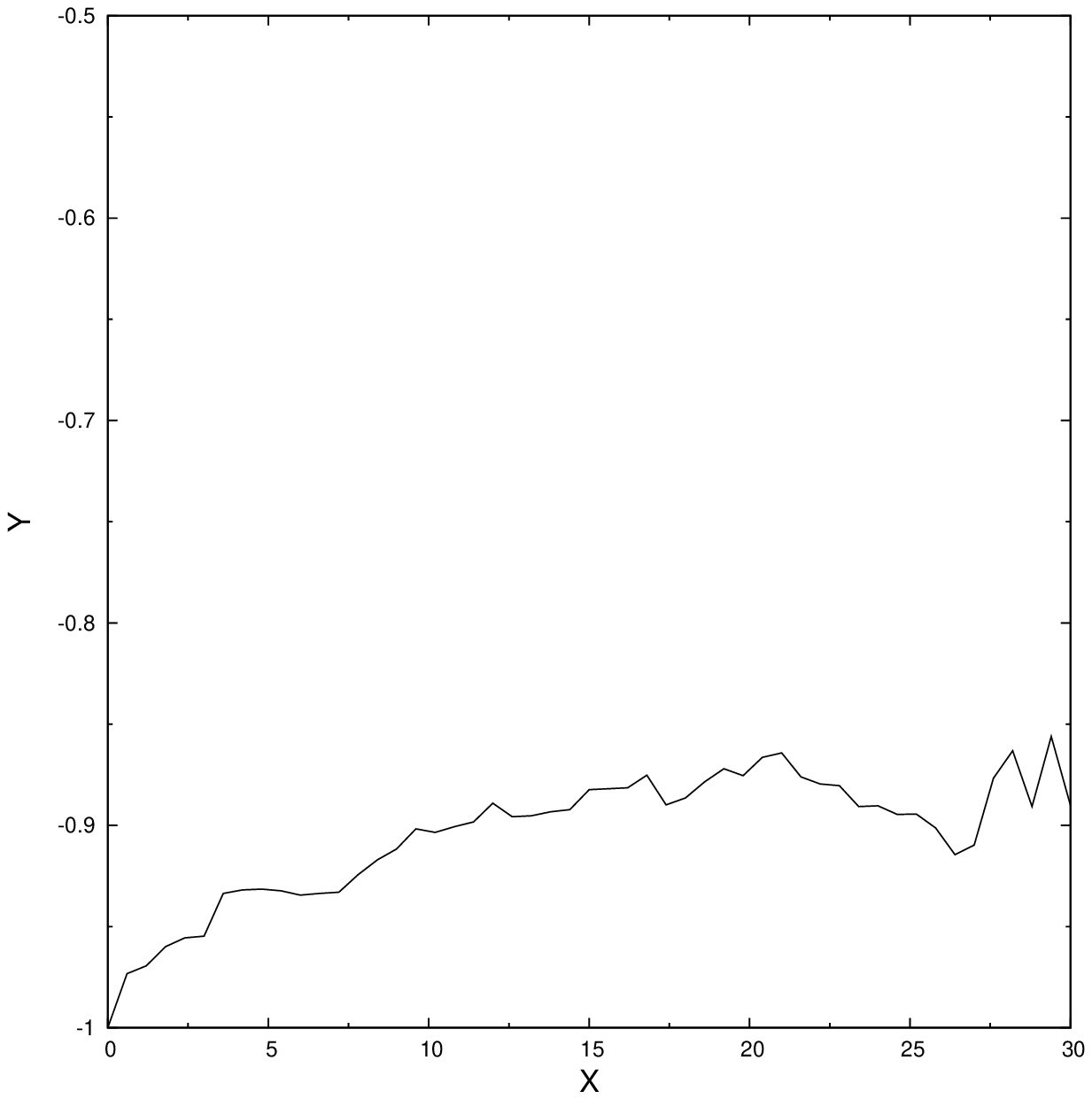}
	\end{center}
	\caption{Dark Matter Radial Density Profile}{The radially averaged dark matter density profile of the large central void in terms of overdensity.}	
\label{fig:void_profile}
\end{figure}

Fig.~\ref{fig:void_profile} shows that the void found by the structure finder indeed replicates a
similar density profile as shown in Fig.~3 of Ref.~\cite{colberg08}.  The void is significantly
under-dense with respect to the overall volume. This reassuringly shows that the
void is truly an under-dense region as expected and verifies that the void finder
works correctly.

\subsection{Feedback Simulation}
\label{sec:gadget_simulation}
Now that it has been shown that the structure finder can adequately find voids, we will explore how well it can self-consistently identify clusters, filaments and voids.  
A simulation was run using the smoothed particle hydrodynamics code,
GADGET-2 \citep{springel05c}, to which we have added the relevant physical processes: radiative
cooling, UV-background heating, star formation, supernovae feedback and metal
enrichment. The radiative cooling and UV-background heating are included by
interpolating the change in entropy from a 4 dimensional table. The table was generated as a 
function of metallicity, redshift, density and temperature using CLOUDY version 10.0, last described in Ref.~\cite{ferland98}.  

The star formation is modeled stochastically by converting the appropriate gas particles into star particles using the recipe described by Refs.~\cite{katz92, kobayashi04}.  
%
Energy and associated metals from supernova are distributed using the prescription
from Ref.~\citep{stinson06}.  The metal yields for core-collapse and Type Ia supernovae are taken
from Ref.~\cite{maoz12a} and  Ref.~\cite{nomoto97b}, respectively.  

Utilizing the 7-year WMAP \citep{komatsu11} results, a $\Lambda$CDM  cosmology was chosen with $\Omega_{m}$ = 0.274, $\Omega_{\Lambda}$ = 0.726, $\Omega_{b}$ = 0.0456 and $h$ = 0.702.  The initial conditions were generated using second-order Lagrangian perturbation theory described by  Ref.~\cite{scoccimarro12}.  The volume investigated was (100 $h^{-1}$ Mpc)$^{3}$ comoving and contained 256$^{3}$ dark matter and 256$^{3}$ gas particles.  

To search for structures, the gas particle density from each snapshot was converted
to a grid. The density at each grid intersection point was computed using
the standard SPH kernel. The density for each volume element (i.e. voxel) in the grid was calculated using the integration of the trilinear interpolating function
constructed from the grid intersection points. This has the advantage of not artificially
washing out features from non-physical smoothing.

\begin{figure}[h]
	\begin{center}
	$\begin{array}{cc}
	\begin{subfigure}{.5 \textwidth}
		\centering
		\includegraphics[width=2.0in]{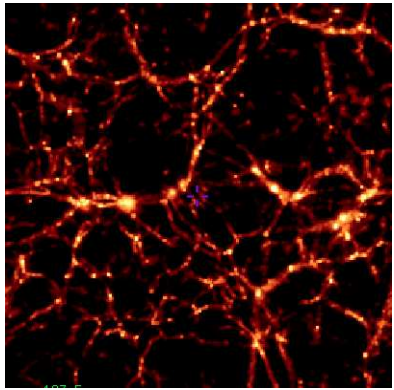}
		\caption{Unfiltered Gas Distribution.}
		\label{fig:gadget_unfiltered}
	\end{subfigure}%
	\begin{subfigure}{.5 \textwidth}
		\centering
		\includegraphics[width=2.0in]{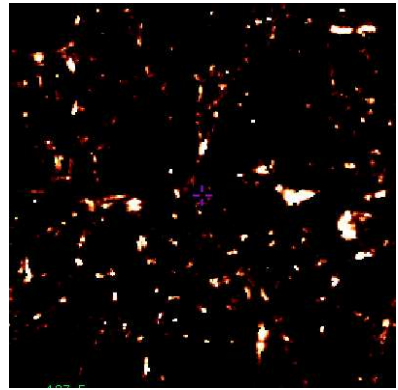}
		\caption{Cluster regions highlighted.}
		\label{fig:gadget_cluster}
	\end{subfigure}\\
	\begin{subfigure}{.5 \textwidth}
		\centering
		\includegraphics[width=2.0in]{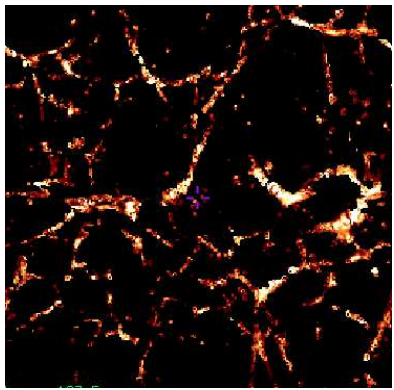}
		\caption{Filament regions highlighted.}
		\label{fig:gadget_filament}
	\end{subfigure}
	\begin{subfigure}{.5 \textwidth}
		\centering
		\includegraphics[width=2.0in]{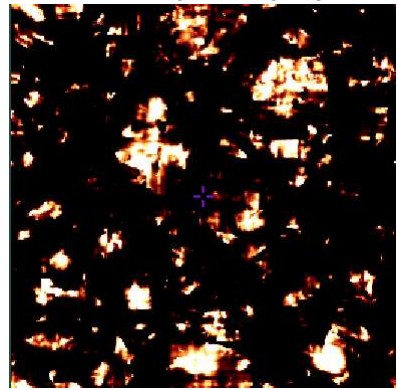}
		\caption{Void regions highlighted.}
		\label{fig:gadget_void}
	\end{subfigure}	
	\end{array}$
	\end{center}
\caption{GADGET-2 Simulation Filtering}{A one voxel thick slice through a selection of a region within a modified GADGET-2 simulation.
Fig.~\ref{fig:gadget_unfiltered} shows the unfiltered gas density log space. Fig.~\ref{fig:gadget_cluster}, Fig.~\ref{fig:gadget_filament} and Fig.~\ref{fig:gadget_void} show filtering with respect to clusters, filaments and voids respectively.
The filtered images appear choppy because we are trying to represent a three
dimensional structure on a two dimensional medium.}
\label{fig:filtering}
\end{figure}

There is no \emph{a priori} knowledge of the best grid resolution to use when converting the particle data to a grid. The 256$^3$ initial gas particles from the
simulation were converted to 128$^3$, 256$^3$ and 512$^3$ grids. Each grid was filtered
with respect to clusters, filaments and voids. The scales that were probed (i.e. $\sigma$
in Eq. \ref{eq:m43}) are listed in  Table \ref{table:2} for the 256$^3$ grid and the physical size of the scales
was maintained for the 128$^3$ and 512$^3$ grids.  Fig.~\ref{fig:filtering} gives a visual example of the filtering.  Qualitatively, the clusters, filaments and voids are found.

After filtering, the structure measures were compared to each other and the structure with the largest measure was selected.  If maximum structure measure was not above a threshold of 0.1 (where the structure measure is normalized on a scale of 0 to 1), no structure type was assigned.  This prevents spurious noise from contributing to the structures.

To determine which grid size to use for the 256$^3$ simulation, the cluster gas fraction from each of the different grid resolutions was compared to the observed gas fraction of the intracluster medium (ICM).  In Ref.~\cite{shull12} the ICM mass fraction
with respect to the baryons is reported to be 4 $\pm$ 1.5 \%.  The 256$^3$ and 512$^3$ grid (see Table \ref{table:3}) agree with this to  within the uncertainty.  This indicates that the Eulerian grid
corresponding to the simulation should have at least the same number of cells as
the SPH simulation has particles.

\begin{figure}[h]
	\begin{center}
	$\begin{array}{cc}
	\begin{subfigure}{.5 \textwidth}
		\centering
		\resizebox{1.0\textwidth}{!}{\input{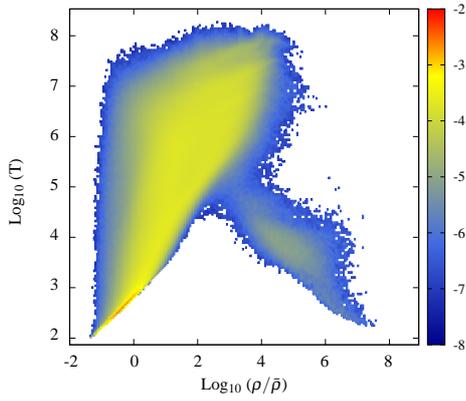}}
		\caption{Total Gas Distribution.}
		\label{fig:total_rho_vs_t}
	\end{subfigure}%
	\begin{subfigure}{.5 \textwidth}
		\centering
		\resizebox{1.0\textwidth}{!}{\input{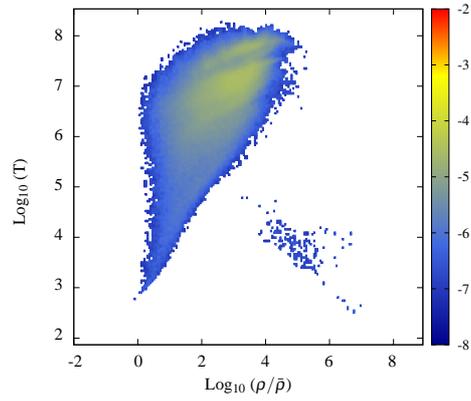}}
		\caption{Clusters.}
		\label{fig:cluster_rho_vs_t}
	\end{subfigure}\\
	\begin{subfigure}{.5 \textwidth}
		\centering
		\resizebox{1.0\textwidth}{!}{\input{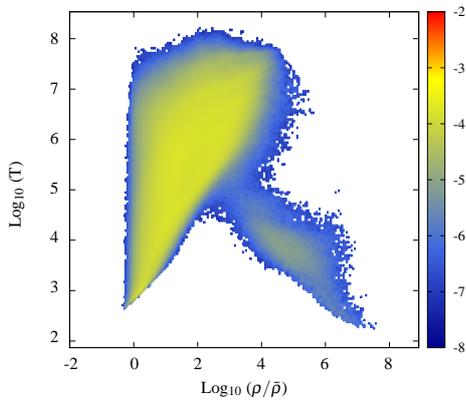}}
		\caption{Filaments.}
		\label{fig:filament_rho_vs_t}
	\end{subfigure}
	\begin{subfigure}{.5 \textwidth}
		\centering
		\resizebox{1.0\textwidth}{!}{\input{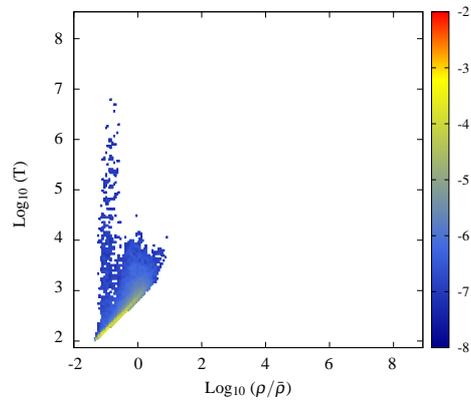}}
		\caption{Voids.}
		\label{fig:void_rho_vs_t}
	\end{subfigure}	
	\end{array}$
	\end{center}
\caption{Density vs. Temperature}{The gas phase distribution in the GADGET-2 simulation as a function of structure type at a redshift of 0.  The color scale gives  the log$_{10}$ of fractional mass at a particular density and temperature.}
\label{fig:rho_vs_t}
\end{figure}

The density vs.~temperature relation was also investigated with respect to
the various structure types shown on the four panels of Fig.~\ref{fig:rho_vs_t}. The different features in the plots are easy to explain. The cluster gas is at high densities and temperatures, as would be expected from a hot intracluster medium.  The void gas is cold and under-dense, as would be expected from regions that lack sufficient density and galaxy interactions to shock heat the gas.  The filaments act as an intermediate link
between the void and cluster regimes in phase space. It is likely that the bulk
of star formation is occurring in the cold, high density regions of the filaments,
but this will be investigated in future work. This verifies that, indeed, the
structure finding algorithm is working as expected.  

\begin{figure}[h]
	\begin{center}
		\psfrag{X}[cc][cc][0.8]{Cluster Measure}
		\psfrag{Y}[cc][cc][0.8]{Filament Measure}
		\includegraphics[width=4.0in]{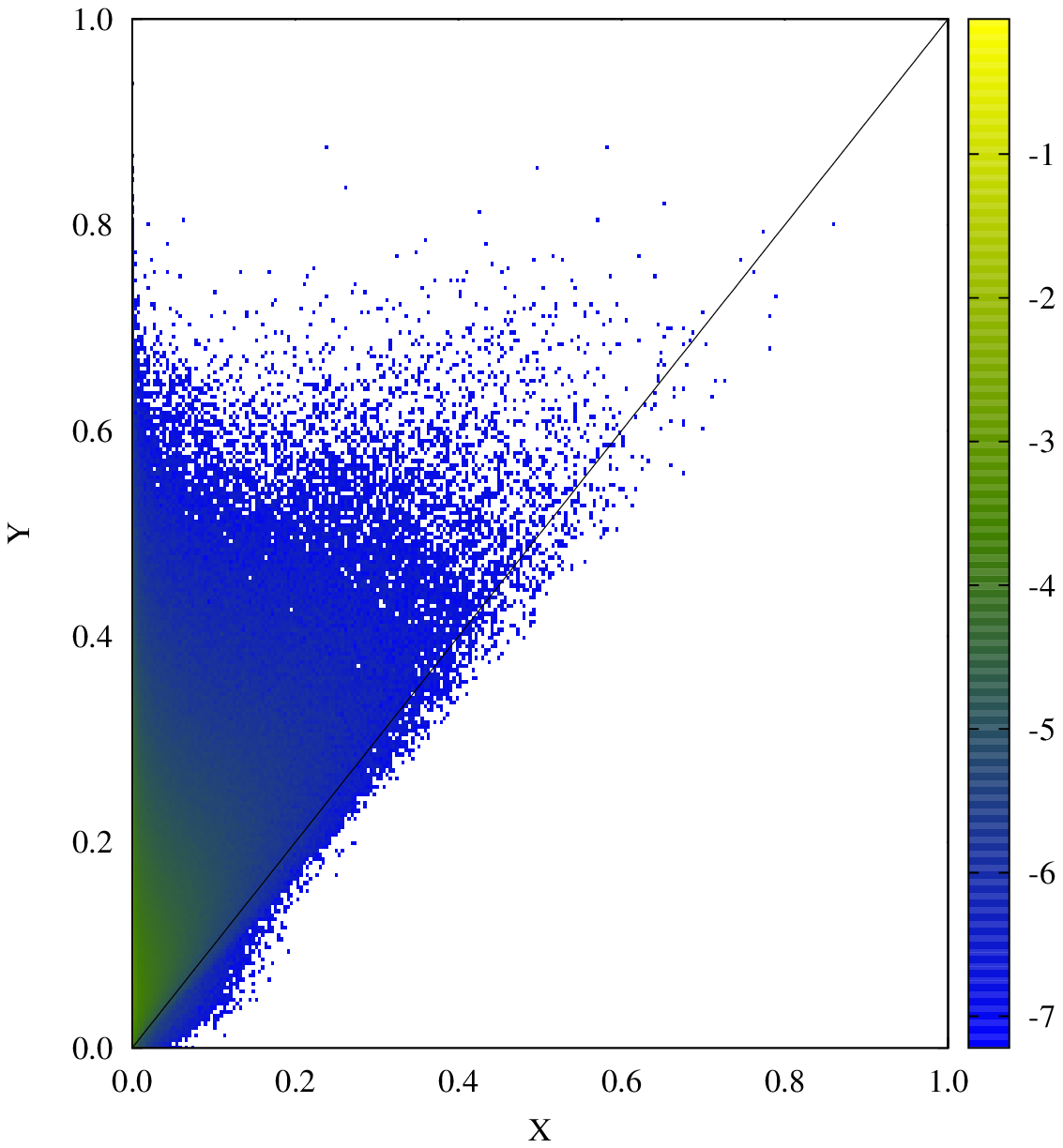}
	\end{center}
	\caption{Cluster vs. Filament Structure Measure}{Comparison of the voxels' filament measure to their cluster measure. Points below the line are classified as clusters while points above the line are classified as filaments. The color bar plots indicates  the $\log_{10}$ of the fraction of voxels at a particular location in the structure-measure space.}
\label{fig:structure_measure}
\end{figure}

Fig.~\ref{fig:structure_measure} shows a comparison of the voxels structure measure from the GADGET-2 simulation (the SPH data has been converted to grid data).  The overlap between filaments and clusters is evident. It originates from regions exhibiting more filamentary structure at certain scales, while at other scales, it is more cluster like. This is why it is necessary to compare the relative values of the structure measure to discern the structural properties that each voxel  exhibits most prominently.  

A similar comparison between filaments and voids yields no overlap.  This is because we have defined voids as being in under-dense regions and filaments in over-dense regions.

\begin{table}[h]
	\centering
	 \begin{adjustwidth}{-1.5cm}{}
	\begin{tabular}{| l | c| c |}
	\hline
		 type  & voxels & $h^{-1}$Mpc \\
		\hline 
		cluster      &    2, 4, 6, 8, 10, 12, 14, 16, 18, 20  			&     0.78, 1.6, 2.3, 3.1, 3.9, 4.7, 5.5, 6.3, 7.0, 8.8\\
		filament   &    1, 2, 3, 4, 5, 6, 8, 10, 12, 14				&     0.39, 0.78, 1.2, 1.6, 1.9, 2.3, 3.1, 3.9, 4.7, 5.5\\
		voids 	   &     5, 7.5, 10, 15, 20, 25, 30, 35, 40, 50	&    1.9, 2.9, 3.9, 5.9, 7.8, 9.8, 11.7, 13.7, 15.6, 19.5 \\
		\hline
		\end{tabular}
		\end{adjustwidth}
	\caption{Scales probed by the structure finding algorithm for the 256$^3$ grid in both voxels and physical space.  The spatial scales were held constant and the voxel scales were adjusted accordingly for the 128$^3$ and 512$^3$ grids.}
\label{table:2}
\end{table}

\begin{table}[h]
	\centering
	\begin{tabular}{| l | c| c |}
	\hline
		 grid  & cluster mass fraction  \\
		\hline 
		128$^3$  &   1.64\% \\
		256$^3$  &   3.16\% \\
		512$^3$  &   2.85\% \\
		\hline
		\end{tabular}
	\caption{Cluster gas mass fraction with respect to the total baryon mass for the 128$^3$, 256$^3$ and 512$^3$ grids.}
\label{table:3}
\end{table}

\section{Discussion}
A self-consistent structure finding algorithm, capable of simultaneously identifying
clusters, filaments and voids, has been presented. The updated algorithm,
has been parallelized using OpenMP and has achieved over a five fold speed up.
The original algorithm was also changed to include the additional void finding
features. It has the advantage over other structure finders \citep{hoyle04, rojas04, ceccarelli06, bond08, colberg08, tinker08, foster09, bond10, kreckel11, platen11, tempel13} in that it provides
a self-consistent method for classifying structures. When compared to
the only other structure finders \citep{aragon07a, hahn07} in the literature capable of finding multiple
types of structures, this algorithm provides simplicity in implementation.

The structure finder can successfully find the same void as the void finders
tested in Ref.~\cite{colberg08}  and has a similar radial density distribution. This verifies
that the void finder is behaving reasonably. When the structure finder was tested
on a large, cosmological simulation it reproduced the expected density vs.~temperature
distribution and the correct fractional amount of the ICM. Future studies
will completely follow the gas phase evolution and the star formation history of
the structures.




\section{Acknowledgments}
\label{sec:acknowledgments}
The subvolume data set of the Millennium simulation was generously provided
by Frazer Pearce. This research was supported in part by the Notre Dame
Center for Research Computing. Work also supported in part by DOE grant
DE-FG02-95-ER40934.

\section{References}
\label{sec:results}
\bibliography{/Users/ali/Library/texmf/bibtex/bib/references}



\end{document}